\documentclass[aps,prd,twocolumn,nofootinbib]{revtex4-1}
\usepackage{graphicx}
\usepackage{amssymb,color}
\newcommand{\non}{\nonumber}

\newcommand{\bfr}{{\bf r}}

\newcommand{\ben}{\begin{displaymath}}
\newcommand{\een}{\end{displaymath}}
\newcommand{\be}{\begin{equation}}
\newcommand{\ee}{\end{equation}}
\newcommand{\bea}{\begin{eqnarray}}
\newcommand{\eea}{\end{eqnarray}}

\newcommand{\eq}[1]{Eq.~(\ref{#1})}

\newcommand{\bfq}{{\bf q}}\def\D{\Delta}
\newcommand{\bfk}{{\bf k}}                                                             

\newcommand{\bfR}{{\bf R}}

\newcommand {\boldtau}{\mbox{\boldmath$\tau$}}
\newcommand {\bfepsilon}{\mbox{\boldmath$\epsilon$}}\def\p{\pi}

\newcommand{\bfhk}{{\bf \hat k}}\newcommand{\bfhq}{{\bf \hat q}}\newcommand{\bfhr}{{\bf \hat r}}
\def\g{\gamma}\def\r{\rho}\def\L{\Lambda}\def\a{\alpha}
\begin{document}

\title{\bf  \hskip10cm NT@UW-19-09\\
{Coherent-Nuclear  Pion Photoproduction and Neutron Radii }}%-photopion19paperv5}}

\author{Gerald A. Miller}

\affiliation{ 
%Department of Physics,
University of Washington, Seattle, WA 98195-1560
}

\date{\today}

\begin{abstract}

{ {\bf Background:}   Knowing the difference between the neutron and proton densities of nuclei is a significant topic because of its importance for understanding neutron star structures and cooling mechanisms. The coherent-nuclear photoproduction of pions, $({\g,\pi^0})$, combined with elastic electron scattering, has been suggested to be a very accurate probe of density differences.
{\bf Purpose:} Study the ${(\g,\pi^0})$ reaction mechanism so as to better access the uncertainties involved in extracting the neutron density.
{\bf Methods:} Include the effects of final-state pion-nucleus charge-exchange reactions on the cross section and study the influence of the non-zero spatial extent of the proton.
{\bf Results:} The effects of final-state charge-exchange increase the cross section between  6\% and  5\%,  generally decreasing as the momentum transfer increases. This leads to an increase of the extracted neutron skin distance by about 50\%. The validity of the previous treatments of the proton size is confirmed.
{\bf Conclusion:} The model dependence of the theoretically computed cross section increases the  total systematic uncertainty (experiment plus theory) in extracting the neutron skin   from the ${(\g,\pi^0})$ cross section  by  at least a factor of three.  }
\end{abstract}  
\maketitle     
\noindent
\section{Introduction}
A recent letter \cite{Tarbert:2013jze}
``establishes the coherent photoproduction  of $\pi^0$ mesons from $^{208}$Pb as an accurate probe of the nuclear shape, which has sufficient sensitivity to detect and characterize the neutron skin."  The neutron skin is the difference  between the neutron and proton densities. That work uses one specific theoretical model to determine  the  neutron skin thickness  to be   $0.15\pm0.03\,({\rm stat.)^{+0.02}_{-0.03}\,(sys.) \,fm}.$ Ref.  \cite{Tarbert:2013jze}
 also points out that the nature of the 
  neutron skin is important  for understanding neutron star structure and cooling mechanisms~\cite{Steiner:2004fi,Horowitz:2000xj,Zu,Steiner2,rutel}, searches for physics beyond the standard model~\cite{nstarsand,parity}, the nature of 3-body forces in nuclei~\cite{Tsang2,Schwenk}, collective nuclear excitations~\cite{centelles,Carbone,Chen,tamii} and flows in heavy-ion collisions~\cite{Li,Tsang}. A nice summary of the relation between nuclear and neutron star physics is provided in \cite{JPPT}.\\
  
As a result, it is important to  understand the theoretical model dependence in computing  $\p^0$  coherent photoproduction cross section in full detail.  Ref.  \cite{Tarbert:2013jze} analyzed its data using the elegant nuclear reaction theory of Ref.~\cite{Drechsel:1999vh}
 based on  the unitary isobar model \cite{Drechsel:1998hk}  for photoproduction  on a free proton. This model gave good agreement with early, less-precise  data on several different nuclei~\cite{Krusche:2002iq,Krusche:2005jx}.  
Possible systematic errors in the theory were {\it  not} included in the analysis \cite{Tarbert:2013jze} that obtained the   neutron skin. One may worry that any systematic error in the theory might have an effect on the extracted skin that is comparable to, or larger than,  the reported systematic error. It is also worth mentioning that alternate reaction theories exist, see {\it e.g.} \cite{Peters:1998mb}.
 Therefore it is responsible to examine whether or not any possible updates are relevant. In particular, effects that are usually ignorable may become relevant given the reported  \cite{Tarbert:2013jze} extraordinarily high precision of $^{+0.02}_{-0.03}$  fm.\\

 The aim here is only to study the possible systematic effects that enter from uncertainties in the reaction theory. Finding a better way to extract the neutron skin from the reaction is a topic that is beyond the scope of this paper.
 Further, it shall be argued that   a variety of contributions to the  systematic error in the theory make that total systematic uncertainty  much  larger than ones originating from experiments.\\
 
\begin{figure}[h]
\includegraphics[width=6.6cm,height=5.65cm]{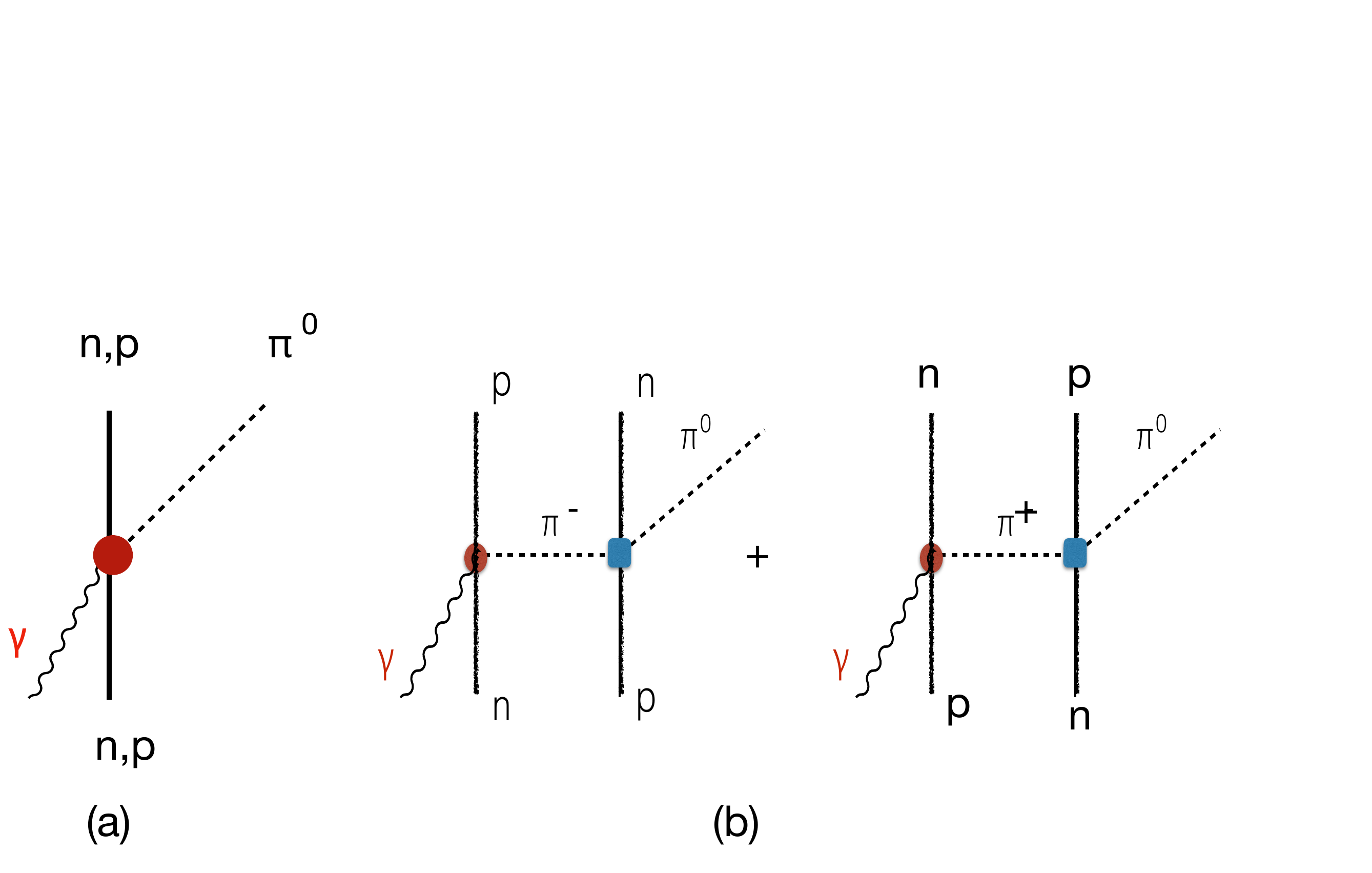}
 \caption{(a) One-body mechanism, impulse approximation. (b) Two-body mechanism, production of charged pion followed by charge-exchange on a second nucleon.}\label{fig:diag}\end{figure}

  The present focus is on the energy range of photon energies between 180 and 190 MeV where the effects of the pion-nucleus optical potential are very small so that the outgoing pions can be treated in plane-wave approximation. 
Even so, in this energy  range  the $(\g,\pi^0)$ reaction proceeds through the excitation of an intermediate $\D$ \cite{Tarbert:2013jze,Krusche:2002iq,Krusche:2005jx}.
 The present note is concerned with two  effects that might lead to a systematic error in the theory. The first  is the effect of production of a charged pion followed by a final state charge exchange reaction leading to the production of a $\p^0$ while leaving the nucleus in its ground state. The relevant diagrams are shown in Fig.~\ref{fig:diag}.  The impulse approximation of Fig.~1(a)  is discussed in Sect.~II.\\

 As explained in the textbook by Ericson \& Weise \cite{Ericson:1988gk}, the processes of Fig. 1b are  dominant for photoproduction near threshold~\cite{Faldt:1979fs,Argan:1981zz}. Moreover,  Wilhelm and Arenh\"ovel \cite{Wilhelm:1996ed} studied the effect of final state charge exchange in the region of the Delta $(\D)$ resonance and found that it causes a significant increase in the computed cross section. 
 Therefore it is necessary to examine the effects of charge exchange. These are  not included in the pion-nucleus optical potential \cite{Gmitro:1987un} used in  Ref.~\cite{Drechsel:1999vh}. However, final state pion-nucleon charge exchange is part of the model  \cite{Drechsel:1998hk} for $\p^0$ production on a nucleon~\cite{Drechsel:1992pn}, and   therefore must be included in the nuclear calculation.This effect is discussed in Sect.~III.\\
 
  The influence  of the non-zero spatial extent of the proton, a subject of much current interest,  is the   second effect examined here. See the reviews \cite{Pohl:2013yb,Carlson:2015jba}.  The radius of the proton is much larger than 0.03 fm, so this effect bears close scrutiny as has been pointed out already in Ref.~\cite{Jones:2014aoa}. This effect is discussed in Sect.~IV. \\

Sect.~V is concerned with the numerical results, and a summary/discussuib is presented in Sect.~VI.
  \section{One-body mechanism}

The dominant one-body (impulse approximation) term is shown in Fig. 1a. The spin-averaged   amplitude for production on a single nucleon, in the notation of  Ericson \& Weise \cite{Ericson:1988gk}, is given by
  \bea &{\cal O}^{IA} =   %{2A^{(3/2)}\over3}\,{2\over3} {\bf\hat q}_{cN}\cdot({\bf \hat k}_{cN}\times \bfepsilon),
{4A^{(3/2)}\over9}\, {\bf\hat q}_{cN}\cdot({\bf \hat k}_{cN}\times \bfepsilon),
 \eea
    in which the incoming photon has momentum $k$,  and the outgoing $\pi^0$ has momentum $q$ in the photon-nucleus center-of-mass (CM) frame.The subscript $cN$ denotes evaluation in the photon-nucleon CM frame
  with the 
  transformation from the lab frame given in  Ref.~\cite{Drechsel:1999vh}. 
The photon transverse polarization is denoted by
   $\bfepsilon$, the photoproduction amplitude for the $\D$ mechanism is written as $A^{(3/2)}$ and the spin-flip term is ignored because the the $^{208}$Pb ground state has no spin.   %The two expressions are the same on-shell but are different in the two  body mechanism
 The resulting nuclear amplitude is given by \bea \langle A|{\cal O}^{IA}|A\rangle= {4A ^{(3/2)}\over9}\, \bfhq_{cN}\cdot(\bfhk_{cN}\times \bfepsilon)\non\\
\times\int d^3r (\rho_n(\bfr)+\rho_p(\bfr))e^{i(\bfk-\bfq)\cdot\bfr},\label{imp}
 \eea where $\bfk$ is the photon momentum in the lab, $\bfk$ in the pion momentum, and  the neutron (n) and proton(p)  densities are given by $\r_{n,p}(\bfr)$. It  is worthwhile to display the explicit forms~\cite{Gmitro:1987un,Chumbalov:1987js} used in \cite{Tarbert:2013jze}:
 \bea& \r(r)=\r_0 {\sinh(c/a)\over \cosh(c/a)+\cosh(c/a)},\label{rhosf}\eea 
 with $ \r_0={3\over 4\pi c^3}{1\over (1+({\pi b\over c})^2)}$, 
   $c$ is the radius parameter and $b$  represents the diffuseness. 
 This density, denoted as the symmetrized Fermi (SF) distribution,  is normalized to unity. It differs from the usual  Fermi function in that the exponential factor in the denominator   is replaced by  the  hyperbolic cosine, and the two forms are identical    in the limit that $b$ goes to 0. This SF  form allows an analytic Fourier transform so that the form factor, $F(q)=\int d^3r \r(r)e^{-i\bfq\cdot\bfr},$ is given by
 \bea F(q)=  {4\pi ^2bc\r_0\over q\sinh(\pi b q)}[{\pi b\over c}\coth(\pi b q) \sin (q c)-\cos(qc)].\non\\\label{fsf}\eea 
 Ref.\cite{Tarbert:2013jze} used $c_p=6.68 $ fm and  $a_p=0.447 $ fm, and extracted  $c_n=6.70$ fm and $ a_n=0.55 $ fm. The form factors using $c_{p,n},\,a_{p,n}$ are denoted as $F_{n,p}(q)$.
 \\
   
      \section{Final State Charge Exchange }
This Section  examines the effect  of charged pion production on one nucleon followed by a pion-nucleon charge exchange reaction, via an s-wave interaction, on a second nucleon. See Fig.~1b.\\

Computation requires knowledge of the photoproduction and the pion-nucleon scattering amplitudes.
The amplitudes for $\gamma N\rightarrow \pi N$ have the general isospin structure
$ A=A^+\delta_{b3}+ A^-{1\over2}[\tau_b,\tau_3]+A^0\tau_b.$
Only  effects of the (3,3) resonance are included, and 
 $A^{(3/2)}=A^+-A^-.$
Only $A^-$ contributes to producing charged pions. This means the  amplitude  $A^-$ is given by 
$ A^-=-A^{(3/2)} .$
The isospin structure of the $\pi N$ scattering system is given by 
 $ T_{ba}=T^+\delta_{ba}+{1\over2}[\tau_b,\tau_a]T^-,$ 
with only the term $T^-$ (giving charge exchange) relevant here.
Given these amplitudes as inputs,  the diagrams of Fig.~\ref{fig:diag} may be evaluated.
\subsection{S-wave final state charge exchange}

 The two body operator ${\cal O}^S_{ji}$ for a pion made on nucleon $i$ to charge-exchange via the S-wave on another nucleon $j$ is  given by
 \begin{widetext}
 \bea 
{\cal O}^S_{ji}=-{A^{(3/2)}\over3}T^- e^{-i\bfq\cdot\bfr_j} 4\pi\int {d^3q'\over(2\pi)^3}{e^{i\bfq'\cdot(\bfr_j-\bfr_i})\over {q'}^2-{q}^2+i\epsilon}I_F(j,i){2\over3}\bfhq'_{cN}\cdot(\bfhk_{cN}\times\bfepsilon)e^{i\bfk\cdot\bfr_i}
,\eea
\end{widetext}
with $ I_F(j,i)=\boldtau_i\cdot\boldtau_j-\tau_3(j)\tau_3(j).$\\

For the kinematics used here  $\bfhq'_{cN}\approx \bfhq'.$  This simplifies the expression so that the integral 
   ${ J}$ given by
\bea&
J=4\pi\int {d^3q'\over(2\pi)^3}{e^{i\bfq'\cdot(\bfr_j-\bfr_i})\over q^2-{q'}^2+i\epsilon}\bfhq'\cdot(\bfhk_{cN}\times\bfepsilon) \eea is relevant.
The use of rotational invariance shows that
\bea
&J= \bfhr_{ji}\cdot(\bfhk\times\bfepsilon) f(r)\eea 
 with   $\bfr_{ji} \equiv \bfr_j-\bfr_i.$
and
\bea
f(r)=4\pi\int {d^3q'\over(2\pi)^3}{e^{i\bfq'\cdot\bfr}\over q^2-{q'}^2+i\epsilon}\bfhq'\cdot\bfhr.\eea
Evaluation of the  integral gives 
\begin{widetext}
\bea& f(r)= %(4\pi)^2 i \int {dq'{q'}^2\over(2\pi)^3} {j_1(q'r)\over q^2-{q'}^2+i\epsilon}\nonumber \\&
 {-2i\over \pi r}+{q  }j_1(qr) +{2i\over \pi}  (\frac{\text{Ci}(q r) (\sin (q r)-q r \cos (q r))-\text{Si}(q r) (q r \sin (q r)+\cos (q r))+q r}{q  r^2}),
%=i\bfnabla_i {e^{iqr_{ij}}\over r_{ij}}\cdot(\bfk\times\bfepsilon)
\eea
\end{widetext}
with $\rm Si$ and $\rm Ci$ being the standard Sine and Cosine integral functions.
  Putting everything together gives the resulting two-nucleon operator: % {\bf this is not yet changed}
\bea
{\cal O}^S_{ji}={2A^{(3/2)}\over9}%{2\over3}
T^- e^{-i\bfq\cdot\bfr_j} I_F(j,i)  f (r_{ji})\widehat{\bfr}_{ji}\cdot(\bfhk\times\bfepsilon)e^{i\bfk\cdot\bfr_i}.
%\equiv\widetilde{ {\cal O}}_{ji} I_F(j,i)
\non\\\eea 
 \subsection{P-wave final state charge exchange}
  
 There is a zero in forward-charge exchange on a nucleon that occurs at  pion kinetic energies of about 50 MeV~\cite{Fitzgerald:1986fg,Irom:1985ni}. This corresponds to the kinetic energy of the pion produced by photons of energies of about 200 MeV.
The amplitude  $T^-$ includes a P-wave term that can be expressed as 
\bea T^-_P=-T^- {\bfhq\cdot\bfhq' },\label{pion}\eea
valid for values of $q$ corresponding to the relevant pion kinetic energies.  Then $T^-+T^-_P=0$ for forward scattering at the appropriate energy.  Including this $P$-wave final state charge exchange reaction 
 leads to another $2N$ contribution denoted as  ${\cal O} ^P$, given by the sum over $i,j$ of
 \begin{widetext} 
 \bea 
{\cal O}_{ji}^P=+{A^{(3/2)}\over3}T^- e^{-i\bfq\cdot\bfr_j} {4\pi }\int {d^3q'\over(2\pi)^3}{e^{i\bfq'\cdot(\bfr_j-\bfr_i})\over {q'}^2-{q}^2+i\epsilon}\bfhq\cdot\bfhq'I_F(i,j){2\over3}\bfhq'\cdot(\bfhk\times\bfepsilon)e^{i\bfk\cdot\bfr_i}
\eea
\end{widetext}
Tensor correlations in the spin-0 nucleus can be ignored, so the integral may be simplified by doing  the angle average over $\bfr\equiv \bfr_i-\bfr_j$. Then
 \bea &
{\cal O}_{ji}^P %=+2{A ^{(3/2)}\over27}T^- e^{-i\bfq\cdot\bfr_j} \bfhq\cdot(\bfhk\times \bfepsilon){ }\int {d^3q'\over(2\pi)^3}{4\pi j_0(q'r) \over {q'}^2-{q}^2+i\epsilon}e^{i\bfk\cdot\bfr_i}\\&
={2A^{(3/2)}\over27}T^- e^{-i\bfq\cdot\bfr_j} \bfhq\cdot(\bfhk\times \bfepsilon){e^{i qr}\over r}e^{i\bfk\cdot\bfr_i}.\eea\\

  \subsection{Nuclear matrix element}
The coherent ground-state to ground state matrix element must be evaluated.
Define 
\bea {\cal O}=\sum_{i\ne j}
{\cal O}_{ji}, 
\eea
with ${\cal O}_{ji}={\cal O}^S_{ji}+{\cal O}^P_{ji}$.
Use second-quantization to get the result
\bea\langle A|{\cal O}|A\rangle=2\sum_{\alpha,\beta,{\rm occupied}}\langle \alpha \beta|{\cal O}_{12}\left(|\alpha\beta\rangle-|\beta\alpha\rangle\right). \eea
%The isospin matrix elements are 
%\bea \langle pp|I_F(1,2)|pp\rangle=0= \langle nn|I_F(1,2)|nn\rangle\\
% \langle np|I_F(1,2)|np\rangle=0= \langle pn|I_F(1,2)|pn\rangle\\
 % \langle np|I_F(1,2)|pn\rangle=2= \langle pn|I_F(1,2)|np\rangle
%\eea
Only the exchange term can contribute, as expected from the diagrams,  so that \bea \langle A|{\cal O}|A\rangle=-2\sum_{\alpha,\beta,{\rm occupied,np}}\langle \alpha \beta|\widetilde{{\cal O}}_{12}|\beta\alpha\rangle .\eea
There are two terms because either $\alpha$ or $\beta$ can denote a neutron, with the other being a proton. %But we have already included the sum over charged pions in evaluating $I_F$ 
The net result is %{\bf for the S-wave term} is 
\begin{widetext}
\bea \langle A|{\cal O}^S|A\rangle=-2 {A^{3/2}\over 3}T^-{2\over3}\int d^3r_1 \,d^3r_2 e^{-i\bfq\cdot\bfr_2}\rho_n(\bfr_2,\bfr_1)
\rho_p(\bfr_1,\bfr_2)e^{i\bfk\cdot\bfr_1}f(r_{12})\widehat{\bfr}_{21}\cdot(\bfhk\times \bfepsilon),\eea and
\bea \langle A|{\cal O}^P|A\rangle=-2 {A ^{3/2}\over 3}T^-{2\over9}\int d^3r_1 \,d^3r_2 e^{-i\bfq\cdot\bfr_2}\rho_n(\bfr_2,\bfr_1)
\rho_p(\bfr_1,\bfr_2)e^{i\bfk\cdot\bfr_1}{e^{iqr_{12}}\over r_{12}}{\bfhq} \cdot(\bfhk\times \bfepsilon).\eea
\end{widetext}
 The term $\r_{n(p)}(\bfr_2,\bfr_1) $ is the neutron $n$ or proton ($p)$ density matrix given by
 \bea \r_{n(p)}(\bfr_2,\bfr_1)\equiv%
 \sum_\a c_{n(p)}^\a \phi_\a^*(\bfr_2)\phi_\a(\bfr_1),
 \eea
where  $\a$ represents the given orbital and $ c_{n(p)}^\a$ represents the occupation number.
The density matrices are evaluated using a local density approximation according to  Negele \& Vautherin \cite{Negele:1975zz}.
Defining 
 $\bfR\equiv{1\over2}(\bfr_1+\bfr_2),\,\bfr\equiv \bfr_1-\bfr_2$
one has
\bea \rho_\nu(\bfr_1,\bfr_2)\approx \rho_\nu(R) P_\nu(r),\eea with
 $P_\nu(r) \equiv {3j_1(k_{F\nu}r)\over k_{F\nu}r}$ and 
  $\nu$ refers to $n,p$.
 Then
 \begin{widetext}
\bea \langle A|{\cal O}^S|A\rangle=-2 {A ^{3/2}\over 3}T^-{2\over3}\int d^3R \,d^3r e^{-i(\bfq-\bfk)\cdot\bfR}\rho_n(\bfR)\rho_p(\bfR)
e^{{ }i(\bfq+\bfk)\cdot\bfr/2}f(r)P_n(r)P_p(r)\widehat{\bfr}\cdot(\bfhk\times \bfepsilon)\eea
\end{widetext}
%\bea \langle A|{\cal O_D}|A\rangle=2 {A_D^{3/2}\over 3}T^-{2\over3}\int d^3R \,d^3r e^{-i(\bfq-\bfk)\cdot\bfR}\rho_n(\bfR)\rho_p(\bfR)
%e^{{\color{red}+}i(\bfq+\bfk)\cdot\bfr/2}f(r)P_n(r)P_p(r)(r)\widehat{\bfr}\cdot(\bfhk\times \bfepsilon)\nonumber\\\eea
The angular integral over $\hat{\bfr}$ is handled first using (with ${\bf V}\equiv {1\over2}(\bfq+\bfk))$
%\bea\int d\hat{\bfr} e^{{\color{red}+}i {\bf V}\cdot \bfr}\hat{\bfr}=\widehat{\bf V} X(Vr)\eea
%where $X$ is a function TBD. Take dot product with $\widehat{\bf V} $ and integrate
The necessary integral is given by 
 $\int d\hat{\bfr}\, e^{ i {\bf V}\cdot \bfr}\hat{\bfr}={ }i\widehat{\bf V}4\pi \,j_1(Vr) $.
Then
\begin{widetext}
\bea &\langle A|{\cal O}^S|A\rangle=%({- }i4\pi)2 {A_D^{3/2}\over 3}T^-{2\over3}(1/2)
{- }{2i\pi\over9} {A^{(3/2)}}T^- 
\nonumber\\&\times\int d^3R  e^{-i(\bfq-\bfk)\cdot\bfR}\rho_n(\bfR)\rho_p(\bfR)
\int r^2dr{ j_1(Vr)}f(r)P_n(r)P_p(r){\bfq\over V}\cdot(\bfhk\times \bfepsilon),\label{S}\eea
and
\bea &\langle A|{\cal O}^P|A\rangle=%({- }4\pi)2 {A_D^{3/2}\over 3}T^-{2\over9}
{- }{4i\pi\over27} {A^{(3/2)}}T^- 
\nonumber\\&\times\int d^3R  e^{-i(\bfq-\bfk)\cdot\bfR}\rho_n(\bfR)\rho_p(\bfR)
\int r^2dr\,j_0(V r) P_n(r)P_p(r){e^{iq r}\over r}{\bfhq }\cdot(\bfhk\times \bfepsilon)\label{P}\eea
\end{widetext}

  Both of the two-body amplitudes depend on the  Fourier transform of the product of neutron and proton densities, $F_2(q)=\int d^3r \r^2(r)e^{-i\bfq\cdot\bfr},$ is presented for comparison purposes. It given by:
\begin{widetext}
\bea
F_2(q)=\frac{(2 \pi   {\rho_0}b)^2  \left(\cos (c q) \left(1-\frac{c \coth
   \left(\frac{c}{b}\right)}{b}-\pi  b q \coth (\pi  b q) \right)+\sin (c q) \left(\pi  \coth \left(\frac{c}{b}\right)
   \coth (\pi  b q)-c q\right)\right)}{q\, \text{sinh}(\pi  b q)}
   .\eea
   \end{widetext}

The Fourier transform of the product of the neutron and proton densities can be obtained to better than about a tenth of a percent, for  relevant values of the  momentum transfer, by using the geometric mean of the neutron and proton radius and diffuseness parameters in the above equation. \\

The complete scattering amplitude, ${\cal M}$ is obtained by summing the terms of \eq{imp}, \eq{S}   and \eq{P} so that
\bea {\cal M}= \langle A|{\cal O}^{IA} +   {\cal O}^S +{\cal O}^P|A\rangle.\label{tot}\eea
This amplitude is squared, and with the appropriate factors used to compute the cross section.

 \section{Non-zero extent of the proton}
  Electron scattering  determines the   charge  nuclear charge density. Computing the $(\g,\pi^0)$ cross section requires the input of the point proton charge density.  Ref.~\cite{Tarbert:2013jze}  obtained this density by using 
parameters   from Klos {\it et al.} [13] \cite{Klos:2007is} who  refer to the charge distribution  of Fricke {\it et al.} \cite{Fricke:1995zz}, together with an approximation given by Oset {\it et al.}  \cite{Oset:1989ey} to transform the charge-density parameters to those for point-protons based on taking into account the proton finite-size. This approximation   is applicable only if $q^2R_A^2\ll 1$.  The relevant momentum transfer here is between 0.3 and 0.9 fm$^{-1}$ so that $q^2R_A^2$ ranges between about 5 and 50. This feature was noted by \cite{Jones:2014aoa} who provided two-parameter Fermi (2PF) function fits to experimental charge and point-proton density.  However, the density used in  \cite{Tarbert:2013jze} is a {\it symmetrized} 2PF  function. The symmetrized Fermi density is given by \cite{Chumbalov:1987js,Gmitro:1987un} \eq{rhosf} and form factor  given by \eq{fsf}.\\
 
 The effects of the spatial extent of the proton's charge density are re-assessed here.
 The point-proton form factor is usually taken as the  nuclear  charge form  factor $F_A(q)=\int d^3r e^{i\bfq\cdot\bfr}\r_A(r)$ divided by $G_E(q)$, the Sachs electric form factor of the proton. Thus the point proton form factor is 
\bea F_{\rm pt}(q) ={F_A(q)\over G_E(q)},\label{fae}\eea
with $F_A(q)=N F_n(q)+Z F_p(q).$. Corrections to \eq{fae} are studied in Ref.~\cite{Miller:2019mae}, but are not included here because the entire effect of the proton size is already known to be small.
The dipole parametrization \bea G_E(q^2)={1\over (1+q^2/\L^2)^2 } \eea generally  represents the data very well in the region with $q\le0.9\,{\rm  fm}^{-1}$ which corresponds to $q^2\le 0.04 \,{\rm GeV}^2$. The exact value of $\L$ is currently in dispute~\cite{Pohl:2013yb,Carlson:2015jba}. A value of $\Lambda=3.93 $ fm$^{-1}$ is used here that corresponds to a proton radius of 0.84 fm. \\

The approximation used previously corresponds to obtaining the correct mean-square radius so that the resulting point form factor $\tilde F_{\rm pt}(q) $ is given by
\bea  \tilde F_{\rm pt}(q) =(1-2q^2/\L^2)F_A(q).\label{faa}\eea The  difference between the exact and approximate form factors is  $\D F_{\rm pt}(q)\equiv F_{\rm pt}(q) - \tilde F_{\rm pt}(q) $ and $\D F_{\rm pt}(q)/ F_{\rm pt}(q)$
is displayed in Fig 2.
The very small values obtained validate  the treatment of Ref.~ \cite{Tarbert:2013jze}. The tiny correction $\D F_{\rm pt}(q)$ is ignored in the following treatment.
\begin{figure}[h]
\includegraphics[width=6.1cm,height=4.15cm]{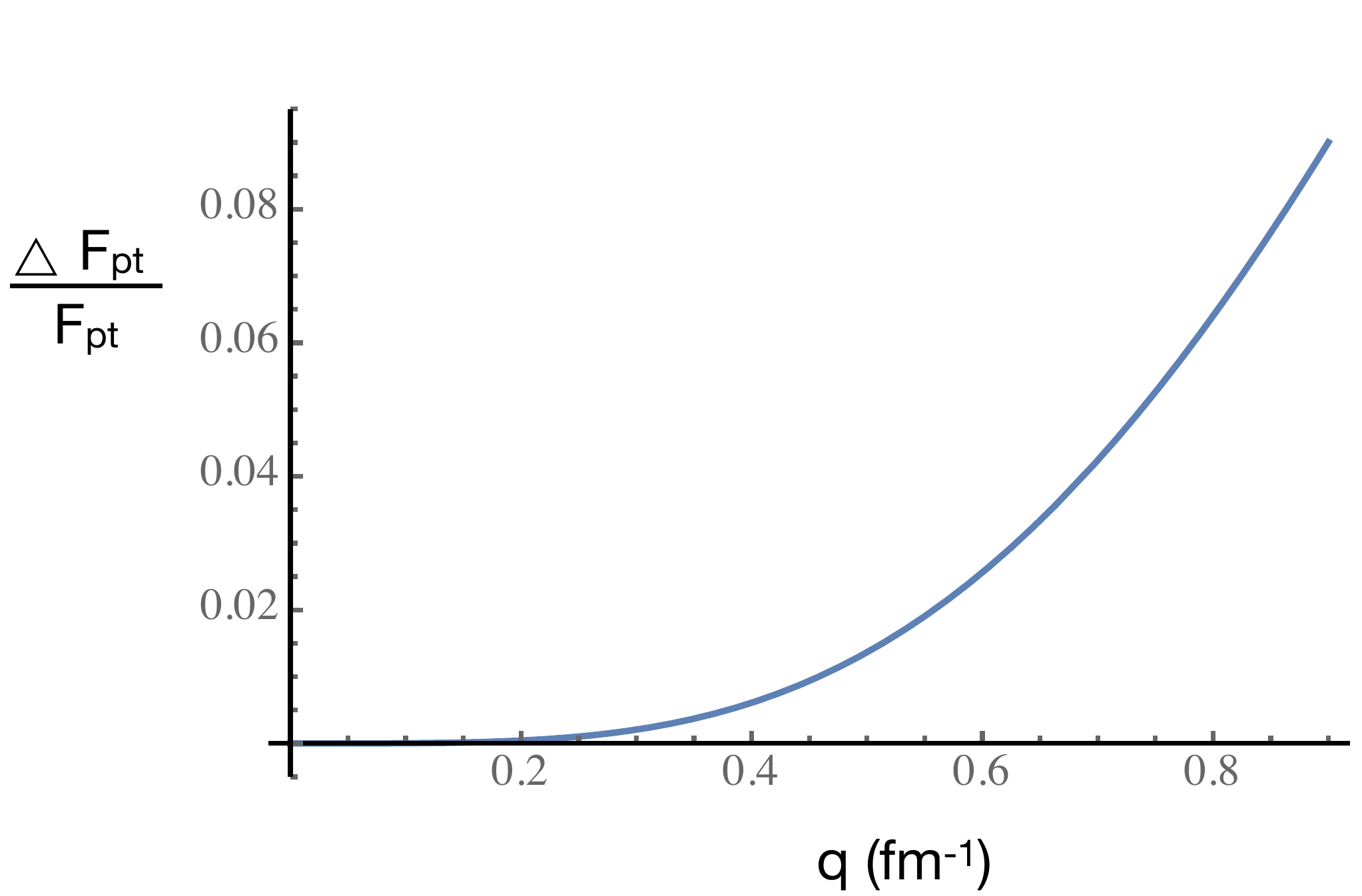}
 \caption{The form factors of \eq{fae}  (solid)and \eq{faa} (dashed).}\label{fig2}\end{figure}

\section{Analysis and Result for the neutron skin}
\begin{figure}[h]
\includegraphics[width=7cm,height=5cm]{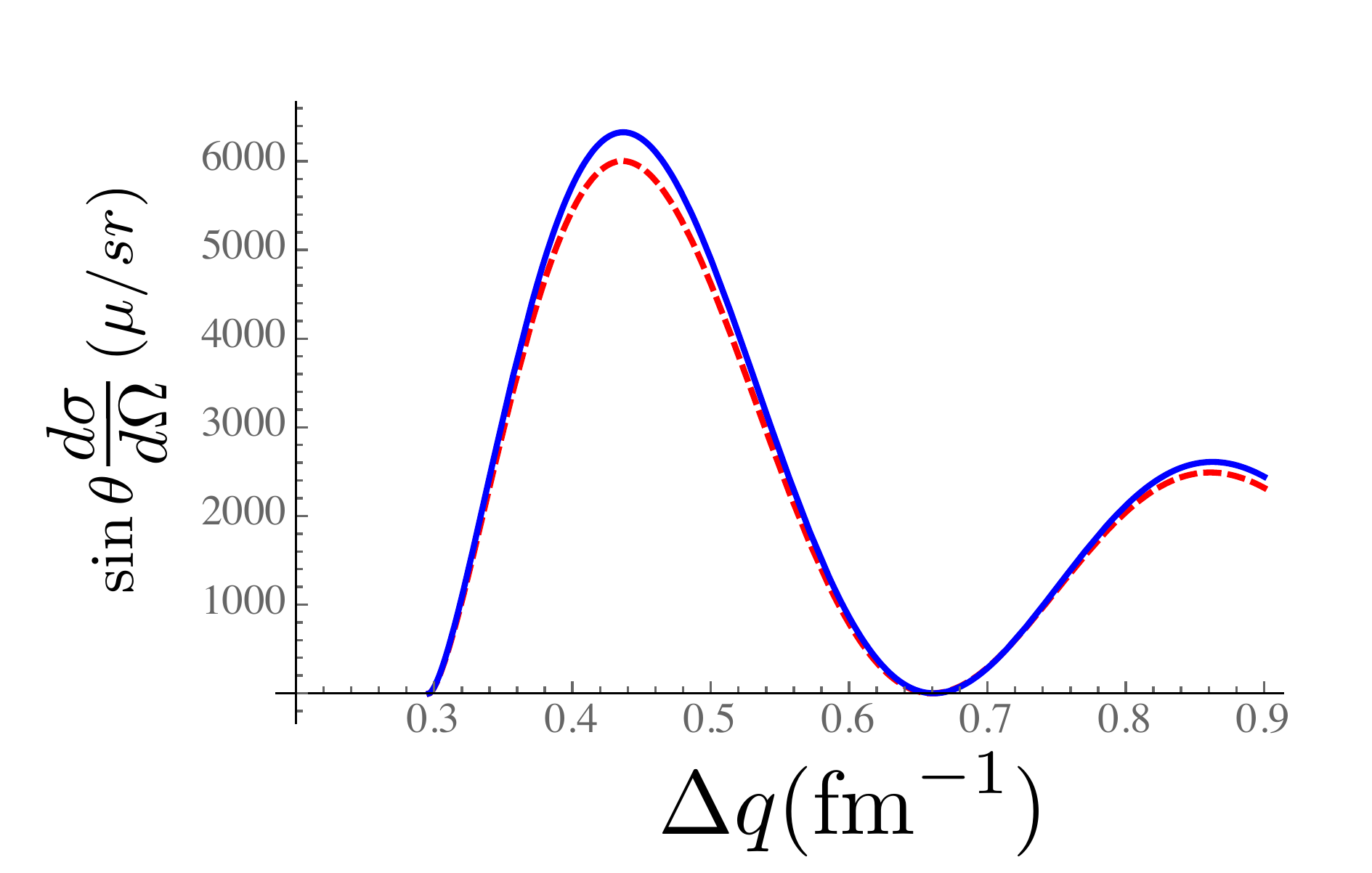}
 \caption{Cross section as a function of momentum transfer $\D q\equiv |\bfk-\bfq|$. Solid (blue) is the complete calculation including the one-body and two body terms. Dashed (red) includes one-body only.}\label{size}\end{figure}
\begin{figure}[h]
\includegraphics[width=7cm,height=5cm]{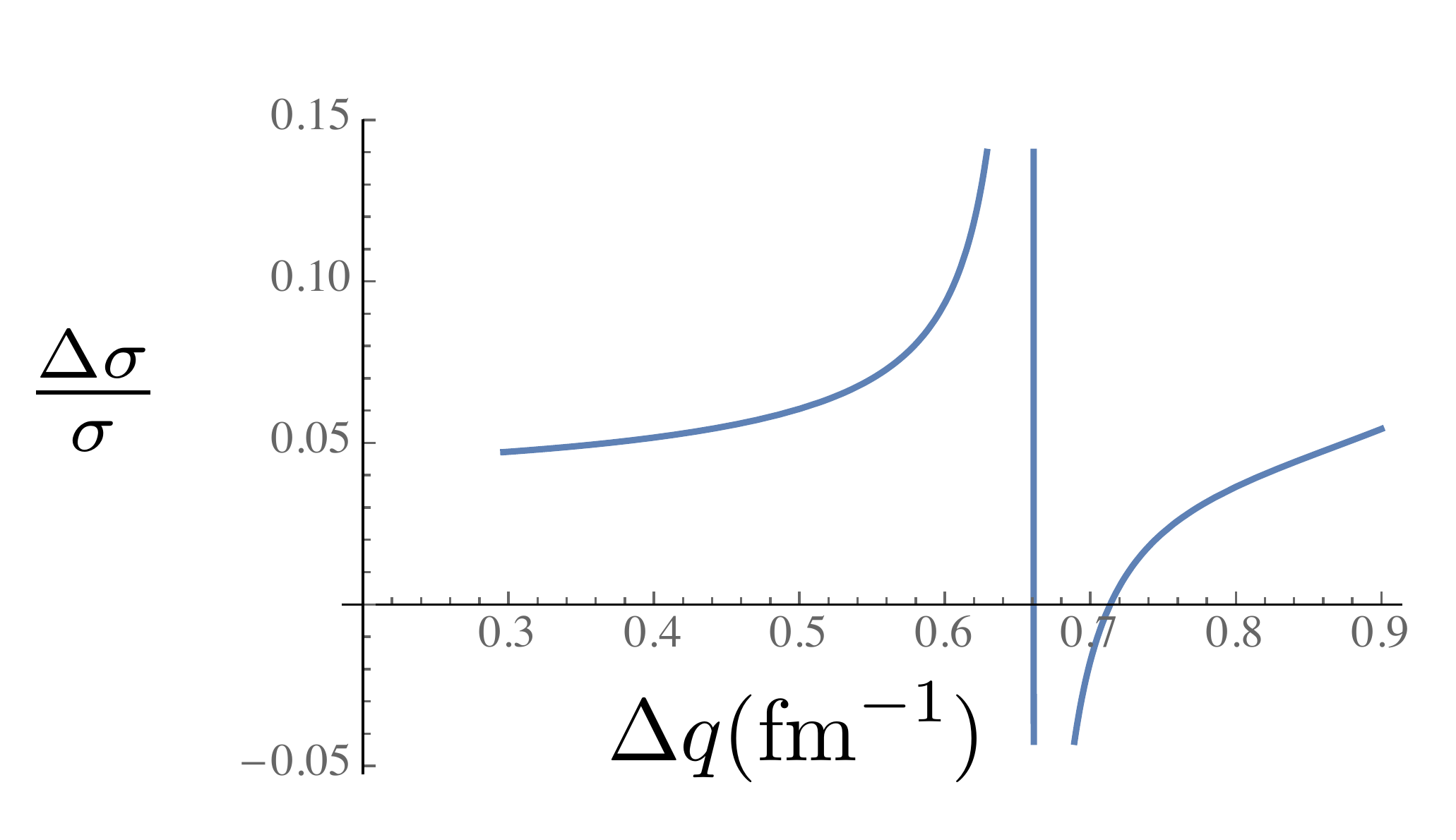}
 \caption{Fractional change of the cross section $\D \sigma$ caused by including the charge exchange final state interaction  as a function of momentum transfer $\D q\equiv |\bfk-\bfq|$.  }\label{pc}\end{figure}

%The neutron has a non-zero charge distribution, with $G_E^n(q^2) $ parametrized as 
%Then the point proton charge density is given by
%\bea &\r_{\rm pt}(r)=\int {d^3q\over (2\pi)^3}{e^{i\bfq\cdot(\bfx-\bfr)}\over G_E(q^2)}\r_A(x)d^3x\\&=
%(1-2 { \nabla^2\over \L^2}+ { \nabla^4\over \L^4})\r_A(r)\eea
The first step is to show the   size of the two-body effects:  Fig.~\ref{size}.
The red dashed curve reproduces the plane wave calculation  shown in  \cite{Tarbert:2013jze}. \footnote{The label of the ordinate axis of Fig.2 of   \cite{Tarbert:2013jze} is missing a factor of $\sin\theta$.} 
  Fig.~\ref{pc} shows the fractional change in the cross section.
The effects of the two body term increase the cross section  by about 6\% at the first maximum and by about 5\% at the second maximum. Including the effects of final-state charge-exchange causes the position of the minimum to be increased by only  about  0.001  fm$^{-1}$. This shift is ignorable, but as a result of this difference, the changes obtain very large magnitudes for values of $\Delta q$ near the minimum.\\

\begin{figure}[h]
\includegraphics[width=7cm,height=5cm]{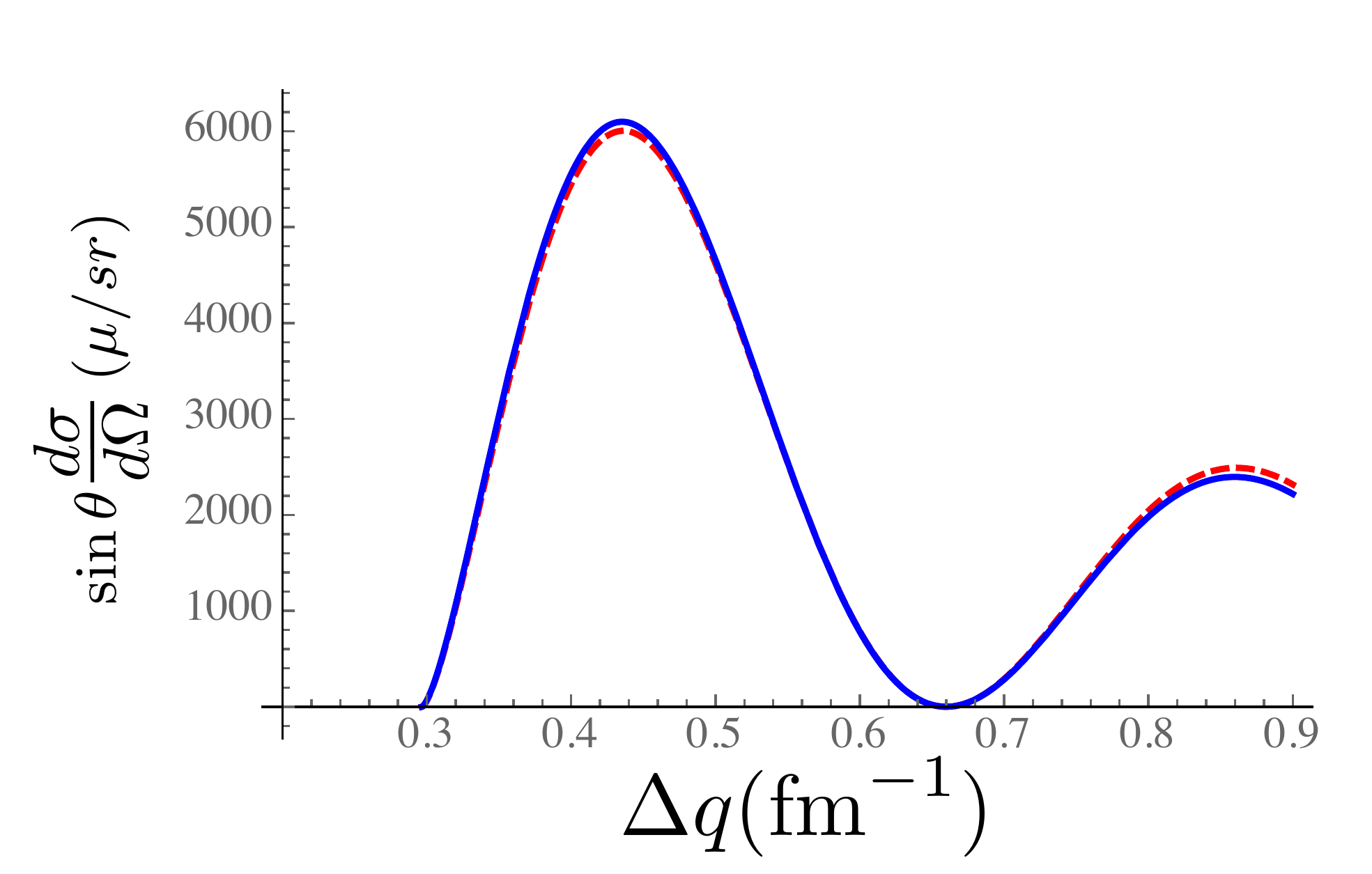}
 \caption{Cross section as a function of momentum transfer $q$. Solid (blue) is the complete calculation including the one-body and two body terms. Dashed (red) includes one-body only.}\label{new1}\end{figure}

The next step is to assess how including the two-body terms of Fig. ~1 impact the extracted value of the neutron skin. To do this,  the cross section obtained from the one-body mechanism with the 
density parameters of   \cite{Tarbert:2013jze} is taken as representing the ``data". Then the complete calculation that includes the charge exchange effect is computed as a function of new values of $a_n$. Values of $a_n$ are varied to find a value that causes the  full calculation (including one- plus two-body amplitudes)  is the same as the ``data". The result is shown in Fig.~\ref{new1}.  Using $a_n=0.61$ fm  instead 0.55 fm in the full calculation leads to a reproduction of the ``data". In Fig.~\ref{new1}    the solid blue (complete calculation)  curve of that overlaps the red dashed (one-body only)  nearly completely. Differences are generally much,  much  larger than the 3\% error assigned to the data in Ref.~ \cite{Tarbert:2013jze}.\\

 Given the new values of $a_n$ and $c_n$ we may compute the neutron skin.  The rms radius $R$ for  a symmetrized Fermi distribution of radius $c$ and diffuseness $a$ is given by the expression:
 \bea R=\sqrt{{1\over5}(3c^2+7\pi^2 a^2)}.\eea
 Using $c_p=6.68$ fm and $a_p=0.447$ gives $R_p=5.43$ fm. Using the values 
 $c_n=6.70$ fm  and $ a_n= 0.55$ fm of Ref.~ \cite{Tarbert:2013jze} gives $R_n=5.58 $ fm, and a skin, \bea \D r_{np}\equiv R_n-R_p,\eea  of $0.143$ fm consistent with the result of that reference. 
 Using instead  $a_n=0.61$ fm and $c_n= 6.7 $ fm which takes the effect of final state charge exchange into account, leads to $R_n=5.79$ fm and a neutron skin of $0.229$ fm. 
 The effects of final state charge exchange are not included in the extraction of the neutron skin  reported by Ref.~ \cite{Tarbert:2013jze}. Including these effects here leads to an increase of the neutron skin by  about  50\%. The same neutron skin  is obtained by increasing the value of  $c_n$ by about 0.10 fm.\\
 
The experimental analysis~\cite{Tarbert:2013jze} did  not use the absolute cross section in extracting the neutron skin, so that the fits in each bin of photon energy have a free normalization parameter~\cite{DW2019}. The theoretical model reproduced the data within ~5-10\% for all bins of the photon energy.
 In the simultaneous fit (diffuseness and half radii) the diffuseness was mainly constrained from the relative heights of the first and second maxima ~\cite{DW2019}. The 5-10\% differences between the theory and experiment are not reflected in the figures in the paper.\\
 
   Changing the normalization to match the data to the theory represents one of the errors in the theory. Here the analogous treatment would be to multiply the first (IA) term of \eq{tot} by the necessary constant needed to reproduce the data.  As a result the  both the theory and the data are represented by the impulse approximation.  Suppose a normalization factor of $\cal N\approx 1$ is needed to match theoretically computed cross section to the data. This means that  the impulse approximation term of \eq{tot} would be multiplied by    $\sqrt{\cal N}$.  Then  the influence of the  final state charge exchange amplitudes  would be  changed by only a factor of $\sqrt{\cal N} -1$.  For example, increasing the computed  cross section by {\it e g.} 5\% to reproduce the  means that the  amplitude ${\cal O}^{IA}$ would be  changed by a factor of about 1.025. The renormalized calculation would then be represented  
 by multiplying the first term ${\cal O}^{IA} $ of \eq{tot} by 1.025. Then  the relative importance of the. charge exchange  terms, ${\cal O}^S+{\cal O}^P$, is reduced only  by 2.5\%. The 6\% increase in the peak cross section   reported above would be changed to an increase of 5.85\%,  The change would truly  be negligible. If ${\cal N}<1$ the importance of the final state charge exchange amplitudes would be {\it increased}. Thus, any uncertainty in  normalization has no impact on the present conclusion that the neutron skin could be 50\% larger than the reported value.   \\

 Moreover, the  theory predicts a significant rise in the cross section as the photon  energy rises from 180 to 240 MeV because the energy approaches that of the $\D$ peak.  The  floating normalization procedure used in Ref.~\cite{Tarbert:2013jze}  loses the opportunity to precisely test the theory.\\

  \section{Summary/Discussion}
  The present effort treats two specific corrections to the reaction mechanism used to extract the neutron density. The effect of charge exchange in the final state leads to a significant (50\%) computed increase in the extracted neutron skin. The effects of the proton's charge density are correctly handled in Ref.~\cite{Tarbert:2013jze}.\\
  
But there are many other uncertainties associated with the pion-nucleus final state interaction that have not been treated here or in   Ref.~\cite{Tarbert:2013jze}. The pion-nucleus optical potential,  necessary to analyze data for photon energies higher than treated here, does not determine pion wave function within the nuclear interior. The resulting ambiguities  have  long been known to lead to significant uncertainties in computing reaction cross sections~\cite{Miller:1974nm,Keister:1978ey}. Moreover, the optical potential    \cite{Gmitro:1987un} used by  Ref.~\cite{Tarbert:2013jze} was constrained only by nuclei with equal numbers of neutrons and protons. In particular, the optical potential  was not tested by comparing to pion-Pb  elastic scattering data.  A key element in the optical potential is the $\D$-nuclear interaction, but no consensus was ever  reached on that interaction \cite{Oset:1979tk,Hirata:1978wp,Freedman:1982yp}.\\

Another issue is that of off-shell effects in the pion-nucleon interaction. The pion-nucleon interaction of \eq{pion} has been instead written as
\bea T^-_P=-\widehat{T}^- {\bfq\cdot\bfq' }\label{new22} \eea because $\widehat{T}^- $ is independent of energy at the low pion energies relevant here. The scattering amplitudes of \eq{pion} and \eq{new22} are the same for on energy-shell kinematic conditions, but differ when $|\bfq'|\ne |\bfq|$. Including this effect increases the amplitude ${\cal O}^P$ by at least 30\%.  A detailed analysis of pion-nucleus elastic scattering data shows that the form of \eq{new22} reproduced all of the systematic features of the data~\cite{Friedman:1983wz}.\\

Other issues involve   potential differences between the reaction theories of ~\cite{Drechsel:1999vh}  and \cite{Peters:1998mb} and the sensitivity of any theory to uncertainties in the $\g-$nucleon interaction that are input to the reaction theory.\\

Treating such problems is far beyond the scope of the present effort, but the discussed previous experience suggests that the related uncertainties are rather large compared to the precision that is relevant for extracting the neutron skin. \\

All of these considerations make it clear that there is a substantial systematic error arising from uncertainties in the theoretical model used to compute the $(\g,\pi^0)$ cross section that was not taken into account in Ref.~\cite{Tarbert:2013jze}.  Given only the size of the effects of the diagrams of Fig. 1b, one can confidently assert that the total (experimental plus theoretical) systematic error was underestimated by at least a factor of three.\\
  
Including the effects of final state charge exchange  along with  the  uncertainties discussed in the present Section suggest that the result for the neutron skin  could be written as  \bea \D r_{np}=0.23\pm0.03\,({\rm stat.)^{+0.02}_{-0.03}\,(sys.) \pm 0.07( th. sys.)\,fm}.\eea  
    \section*{Acknowledgements} This   work has been partially supported by the U. S. Department of Energy Office of Science, Office of Nuclear Physics under Award Number DE-FG02-97ER-41014.
  %  I thank C. J. Horowitz and A. Gardestig for useful discussions.

 \end{document}